\documentclass[12pt]{article}
\usepackage[]{fullpage}

\usepackage{graphicx}                    
\usepackage{natbib}                     
\usepackage{amsmath}
\usepackage{amssymb}                    
\usepackage{color}                       
\usepackage{subfigure}
\usepackage{bbold}
\usepackage{url}     
\usepackage{tabularx}

\usepackage{authblk}

\title{Correlations at large scales and the onset of turbulence in the fast solar wind}%
\author[1]{R. T. Wicks}
\author[1]{D. A. Roberts}
\author[2]{A. Mallet}
\author[2]{A. A. Schekochihin}
\author[3]{T. S. Horbury}
\author[4]{C. H. K. Chen}

\affil[1]{Code 672, NASA Goddard Space Flight Center, Greenbelt, MD, USA }
\affil[2]{Rudolf Peierls Centre for Theoretical Physics, University of Oxford, Oxford, OX1 3NP, UK}
\affil[3]{Space and Atmospheric Physics Group, Imperial College London, London, SW7 2AZ, UK}
\affil[4]{Space Sciences Laboratory, University of California, Berkeley, California 94720, USA}

\begin{document}
\maketitle
\date{\today}
\begin{abstract}
We show that the scaling of structure functions of magnetic and velocity fields in a mostly highly Alfv\'{e}nic fast solar wind stream depends strongly on the joint distribution of the dimensionless measures of cross helicity and residual energy. Already at very low frequencies, fluctuations that are both more balanced (cross helicity $\sim 0$) and equipartitioned (residual energy $\sim 0$) have steep structure functions reminiscent of ``turbulent" scalings usually associated with the inertial range. Fluctuations that are magnetically dominated (residual energy $\sim -1$), and so have closely anti-aligned Elsasser-field vectors, or imbalanced (cross helicity $\sim 1$), and so have closely aligned magnetic and velocity vectors, have wide `$1/f$' ranges typical of fast solar wind. We conclude that the strength of nonlinear interactions of individual fluctuations within a stream, diagnosed by the degree of correlation in direction and magnitude of magnetic and velocity fluctuations, determines the extent of the $1/f$ region observed and thus the onset scale for the turbulent cascade.
\end{abstract}


 
\section{Introduction} 

The solar wind is a continuous supersonic flow of plasma emitted by the Sun. Observations made {\it in-situ} by spacecraft provide long, high-cadence time series well suited to the study of magnetohydrodynamic (MHD) plasma turbulence \citep{Goldstein95}. MHD turbulence mediates the transfer of energy at scales larger than the proton gyroscale, in the ``inertial range" where dissipation of fluid motions is negligible. Turbulent fluctuations scatter energetic particles such as cosmic rays and solar energetic particles as well as providing energy to heat the solar wind.  Given the observational ubiquity of plasma turbulence throughout the universe, results deduced from observations of the solar wind are important in many areas of astrophysics.
\par
The frequency-dependent Fourier spectrum of time-series observations of the solar wind magnetic field ({\bf B}) can be directly related to the wavenumber ($k$) spectrum by the Taylor hypothesis. In the inertial range, the energy spectrum of magnetic fluctuations typically scales as $E(f) \propto f^{-5/3}$. At larger scales this spectrum is often observed to have a distinct low-frequency range with $E(f) \propto f^{-1}$ \citep{Burlaga84, Matthaeus86, Matthaeus07, Smith95}. Solar wind streams containing the most correlated magnetic field and velocity fluctuations, typically found in co-rotating fast streams and fast polar wind, have the widest $1/f$ ranges. Less correlated streams have shorter, or no, $1/f$ range \citep{Goldstein84, Matthaeus07, Tu89, Tu90}. The spectral break between the $1/f$ range and the inertial range is observed to move to lower frequencies with increasing distance from the Sun \citep{Bavassano82, Horbury96, Roberts10} and the fluctuations within the $1/f$ range reduce in amplitude with distance ($R$) from the Sun $\propto R^{-3}$ consistent with the WKB approximation \citep{Roberts89, Roberts90, Jokipii95, Horbury96}. These results support the idea that the energy in the $1/f$ range is contained in linear superpositions of coronal structures and Alfv\'{e}nic fluctuations of solar origin, which do not evolve significantly until they become turbulent, acting as an energy reservoir for the turbulence \citep{Matthaeus86, Hollweg90}. 
\par
However, recent studies have shown that the behavior of fluctuations at large scales in the fast solar wind is more complicated than allowed for by the WKB model. When fluctuations in the fast wind are less correlated, scaling of third-order moments is observed to extend to very low frequencies \citep{SorrisoValvo07, Marino12}, perhaps a sign of active turbulence. Scaling of the alignment angle between {\bf B} and velocity {\bf V} fluctuations in the $1/f$ range of the solar wind has also been observed \citep{Podesta09, Hnat11}, suggestive of some evolution of the nature of the fluctuations with scale. \cite{Wicks13} showed that the angle between oppositely propagating Elsasser fluctuations increases with increasing frequency in the $1/f$ range, with the fluctuations becoming anti-aligned. Furthermore, fluctuations with perpendicular alignment showed steeper scaling than aligned fluctuations, suggesting an active turbulent cascade for the unaligned sub-population of fluctuations in the $1/f$ range. 
\par
Two dimensionless parameters, the normalized residual energy (\ref{eq:NRE}) and the normalized cross helicity (\ref{eq:NCH}) together completely define the two-dimensional geometry of the fluctuations in the plane formed by the two vectors, (see, e.g., Equations 7 and 8 and Figure 4 in \cite{Wicks13}).
\begin{align}
\sigma_r = \frac{\delta\textbf{v}^2 - \delta\textbf{b}^2}{\delta\textbf{v}^2 + \delta\textbf{b}^2}, \label{eq:NRE}\\
\sigma_c = \frac{2\delta\textbf{v}\cdot\delta\textbf{b}}{\delta\textbf{v}^2 + \delta\textbf{b}^2}, \label{eq:NCH}
\end{align}
where the Alfv\'{e}n-normalized magnetic field fluctuation is $\delta\textbf{b} = \delta\textbf{B}/\sqrt{\mu_0\rho}$ (where $\rho$ is the plasma density) and the velocity fluctuation is $\delta\textbf{v}$ (see Equations \ref{eq:db} and \ref{eq:dv} below). The nonlinear terms in the MHD equations are dependent on the geometry of the fluctuations in {\bf B} and {\bf V} relative to one another \citep{Elsasser, DMV1980, Podesta09, Wicks13}, in particular the relative amplitudes of the fluctuations and the angle between them. Thus the strength of the nonlinear interaction may change across the phase space defined by $\sigma_c$ and $\sigma_r$. These two dimensionless parameters are correlated, as shown by \cite{Bavassano98, Bavassano06}, who measured the joint distribution of $\sigma_c$ and $\sigma_r$. Their joint distributions were also shown to change slightly with scale \citep{Bavassano98, Bavassano06} and with the distance from the Sun \citep{DAmicis10}, echoing the theoretical expectation that nonlinear interaction changes the correlations of {\bf B} and {\bf V}.
\par
\begin{figure*}[t]
\includegraphics[width=\textwidth]{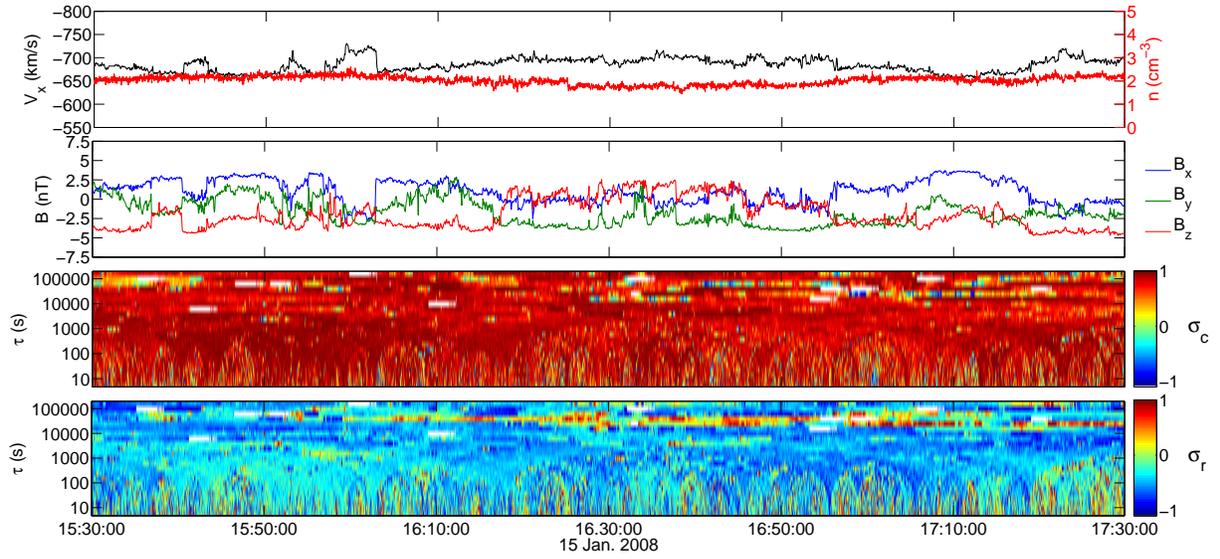}
\caption{Time series at 3~s resolution from two hours of the seven-day-long fast stream used in this analysis. The top panel shows the solar wind speed and density. The second panel shows the three components of the magnetic field vector. The bottom two panels show the normalized cross helicity and residual energy (Equations \ref{eq:SCp}-\ref{eq:SRp}) computed at different time scales $\tau$. White areas correspond to times when a data gap is selected at $t$ or $t+\tau$ in equations \ref{eq:SCp}-\ref{eq:SRp}. There are no data gaps in the two hour window shown but $\tau > 7.2\times10^3$~s will select a point $t+\tau$ outside of the data in the plot.} 
\label{fig:0}
\end{figure*}
Here we combine the approach of \cite{Bavassano98, Bavassano06, DAmicis10} with that of \cite{Wicks13} to investigate the effect of the joint distribution of the local normalized cross helicity and residual energy on the scaling of structure functions of the fluctuating fields. The aim of this study is to look for systematic effects on the width of the $1/f$ range due to the correlation properties of fluctuations within a single solar wind stream. Using a single stream aids the analysis by fixing the external variables that change between different streams: the travel time from the Sun and the evolution of plasma parameters such as plasma $\beta$ and the Alfv\'{e}n speed. This study thus differs from previous studies in the solar wind \citep{Burlaga84, Goldstein84,  Roberts87, Tu89, Tu90} that used the average correlation properties of many streams. 

\section{Data}

We use 3-second Wind spacecraft MFI and 3DP observations of the magnetic field {\bf B}, proton density $\rho$, and velocity {\bf V} taken from a fast solar wind stream from Jan.~14 04:40:00 to Jan.~21 03:20:00 of 2008. The average solar wind conditions were: speed $|V| = 660$~km/s, magnetic field $|B| = 4.4$~nT, proton number density $n_p = 2.4$~cm$^{-3}$, Alfv\'{e}n speed $V_A = 62$~km/s, and the ratio of thermal to magnetic pressure for protons $\beta_p = 1.2$. Three similar co-rotating, approximately seven-day-long fast streams are observed from December 2007 to March 2008 with good data coverage (no individual data gaps longer than 3 hours and total data coverage of 90\% or better). All three give rise to results that are quantitatively similar to those shown here.
\par
Increments in Alfv\'{e}n-normalized $\textbf{B}$ and $\textbf{V}$ are calculated as a function of the time lag $\tau$:
\begin{align}
\delta\textbf{b}(t,\tau) =& \frac{\textbf{B}(t) - \textbf{B}(t+\tau)}{\sqrt{\mu_0\rho_0(t,\tau)}},\label{eq:db}\\
\delta\textbf{v}(t,\tau) =& \textbf{V}(t) - \textbf{V}(t+\tau).\label{eq:dv}
\end{align}
The local mean field $\textbf{B}_0(t,\tau)$ and the local mean proton density $\rho_0(t,\tau)$ are calculated over the same time scales $\tau$:
\begin{align}
\textbf{B}_0(t,\tau) =& \frac{1}{\tau}\int\limits_{t' = t}^{t' = t+\tau}{\textbf{B}(t')}\mathrm{d}t',\label{eq:B0}\\
\rho_0(t,\tau) =& \frac{1}{\tau}\int\limits_{t' = t}^{t' = t+\tau}{\rho(t')}\mathrm{d}t'.\label{eq:r0}
\end{align}
We use the component of the fluctuations perpendicular to the local field in order to select the Alfv\'{e}nic part of the fluctuations, minimizing the effect of compressible and pseudo-Alfv\'{e}nic fluctuations on our results \citep{Wicks12}:
\begin{align}
\delta\textbf{b}_{\perp}(t,\tau) =& \delta\textbf{b}(t,\tau)\cdot\left(\mathbb{1} - \hat{\textbf{b}}_0(t,\tau)\hat{\textbf{b}}_0(t,\tau)\right),\label{eq:dbperp}\\
\delta\textbf{v}_{\perp}(t,\tau) =& \delta\textbf{v}(t,\tau)\cdot\left(\mathbb{1} - \hat{\textbf{b}}_0(t,\tau)\hat{\textbf{b}}_0(t,\tau)\right),\label{eq:dvperp}
\end{align}
where $\hat{\textbf{b}}_0 = \textbf{B}_0/B_0$ and $\mathbb{1}$ is the unit matrix \footnote{Note, however, that the perpendicular Alfv\'{e}nic fluctuations account for around $90\%$ of the energy in the turbulence and using the full vector instead of its perpendicular part does not qualitatively change the results.}. We also calculate Elsasser variables to give an estimate of the imbalance between the outward ($\delta\textbf{z}^{+}$) and inward ($\delta\textbf{z}^{-}$) propagating fluctuations\footnote{There are no magnetic sector boundaries in the time series and so no accounting for changing definitions of ``inward" and ``outward" is required.}:
\begin{align}
\delta\textbf{z}^{\pm}_{\perp}(t,\tau) =& \delta\textbf{v}_{\perp}(t,\tau) \pm \delta\textbf{b}_{\perp}(t,\tau).\label{eq:dz}
\end{align}
The perpendicular fluctuations are used to calculate scale-dependent normalized cross helicity and residual energy: 
\begin{alignat}{2}
\sigma_c(t,\tau) &= \frac{2\delta\textbf{v}_{\perp}(t,\tau)\cdot\delta\textbf{b}_{\perp}(t,\tau)}{|\delta\textbf{v}_{\perp}(t,\tau)|^2+|\delta\textbf{b}_{\perp}(t,\tau)|^2} \label{eq:SCp}\nonumber\\
 &= \frac{|\delta\textbf{z}^+_{\perp}(t,\tau)|^2-|\delta\textbf{z}^-_{\perp}(t,\tau)|^2}{|\delta\textbf{z}^+_{\perp}(t,\tau)|^2+|\delta\textbf{z}^-_{\perp}(t,\tau)|^2},\\
\sigma_r(t,\tau) &= \frac{|\delta\textbf{v}_{\perp}(t,\tau)|^2-|\delta\textbf{b}_{\perp}(t,\tau)|^2}{|\delta\textbf{v}_{\perp}(t,\tau)|^2+|\delta\textbf{b}_{\perp}(t,\tau)|^2}\label{eq:SRp}\nonumber\\
 &= \frac{2\delta\textbf{z}^+_{\perp}(t,\tau)\cdot\delta\textbf{z}^-_{\perp}(t,\tau)}{|\delta\textbf{z}^+_{\perp}(t,\tau)|^2+|\delta\textbf{z}^-_{\perp}(t,\tau)|^2}. 
\end{alignat}
\par
Two hours from the 7-day-long fast stream that we used are shown in Figure \ref{fig:0}. The solar wind speed and density, shown in the top panel, are approximately constant and typical for a fast wind interval at 1 AU. The magnetic field, shown in the second panel, fluctuates with time. The cross helicity and residual energy calculated over a range of scales $\tau$ are shown in the bottom two panels. Typically the cross helicity is positive (red) and the residual energy is negative (blue). This implies that the fluctuations tend to be correlated (magnetic and velocity fluctuations are aligned) but have somewhat larger magnetic-field component than velocity component. The two quantities are correlated, with more positive cross helicity typically coinciding with residual energy close to zero, and more negative residual energy coinciding with cross helicity closer to zero (which is geometrically inevitable, see below).
\par
\begin{figure}[t]
\centering
\includegraphics[width=0.75\textwidth]{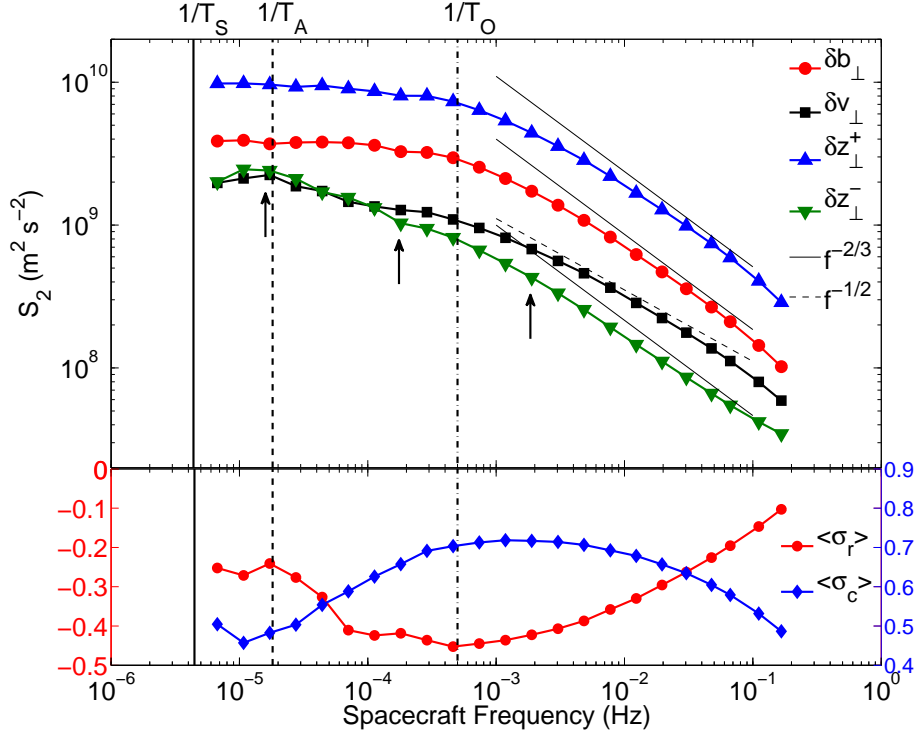}
\caption{Top panel: structure functions of the vector fields are shown plotted against spacecraft frequency ($f = 1/\tau$). Bottom panel: normalized cross helicity and residual energy (Equations \ref{eq:SCp}-\ref{eq:SRp}) averaged over the entire duration of the fast stream. The arrows indicate the lags $\tau$ used in Figure \ref{fig:3}, with the middle arrow also indicating the scale used in Figure \ref{fig:1}.} 
\label{fig:0.1}
\end{figure}
We measure the scale-dependent amplitude of the fluctuations using second-order structure functions 
\begin{align}
S_2(\delta\textbf{b}_{\perp}, \tau) &= \left< \left| \delta\textbf{b}_{\perp}(t,\tau) \right|^2 \right>, \label{eq:S2}
\end{align}
and similarly for $\delta\textbf{v}_{\perp}$ and $\delta\textbf{z}^{\pm}_{\perp}$. The average is done over the entire length of the stream. The mean properties of the fluctuations over the 7-day period are summarized in Figure \ref{fig:0.1}, which shows the structure functions vs.~frequency $f \equiv 1/\tau$ in the top panel and the associated mean $\sigma_c$ and $\sigma_r$ in the bottom panel. 
\par
The two vertical lines at low frequencies mark two important scales. $1/T_S$ is the frequency defined by the time the solar wind takes to flow from the Sun to the spacecraft at 1 AU. $1/T_A$ is the frequency corresponding to the advection past the spacecraft of the largest distance an Alfv\'{e}n wave can have travelled in the time the solar wind has propagated from the Sun to the spacecraft. We estimate the latter by using average values of the solar wind speed and the Alfv\'{e}n speed during the 7-day-long stream and assuming that the Alfv\'{e}n speed changes with distance from the Sun, $R$, as $V_A \propto R^{-1/2}$ (see \cite{Wicks13}). This time-scale is a rough estimate of the upper limit on the time lag over which Alfv\'{e}nic fluctuations may have interacted and so acts as the approximate low-frequency limit below which turbulence cannot develop. 
\par
The structure functions for the four vector fields are plotted individually and conform to the expected behavior of structure functions in highly Alfv\'{e}nic fast solar wind. The structure function scaling exponents, $\alpha$, are related to power spectral indices, $\gamma$, via $\gamma = \alpha-1$ for $0 < \alpha < -2$, so $-2/3$ corresponds to the $-5/3$ ``Kolmogorov" spectral slope. The structure functions of the magnetic field and the outward propagating Elsasser fluctuations $(\delta\textbf{z}^+)$ have extended flat ranges at low frequencies; this is the ``$1/f$" spectral range. Over the range of frequencies where the magnetic-field structure functions are flat, the structure functions of the velocity and the inward propagating Elsasser fluctuations $(\delta\textbf{z}^-)$ have a shallow scaling. The boundary between the energy-containing scales in the $1/f$ range and the turbulent inertial range is estimated as the ``knee" in the magnetic field structure functions where the scaling changes from flat to $f^{-2/3}$. This frequency is indicated in Figure \ref{fig:0.1} and later figures by the vertical line labelled $1/T_O$. We will refer to this frequency as ``the outer scale". At this frequency, all of the four structure functions steepen, those of the magnetic field and both Elsasser variables to logarithmic slopes close to $-2/3$, whereas the velocity structure function has a slope of $-1/2$, corresponding to $-5/3$ and $-3/2$ spectra, respectively \cite[see, for example,][]{Podesta07, Roberts10}. 
\par
The bottom panel in Figure \ref{fig:0.1} shows the mean cross helicity $\left<\sigma_c\right>$ and residual energy $\left<\sigma_r\right>$ calculated from the field increments (Equations \ref{eq:SCp} and \ref{eq:SRp}):
\begin{alignat}{2}
\left<\sigma_c\right> &= \left<\frac{2\delta\textbf{v}_{\perp}(t,\tau)\cdot\delta\textbf{b}_{\perp}(t,\tau)}{|\delta\textbf{v}_{\perp}(t,\tau)|^2+|\delta\textbf{b}_{\perp}(t,\tau)|^2}\right>, \label{eq:SCpAv}\\
\left<\sigma_r\right> &= \left<\frac{|\delta\textbf{v}_{\perp}(t,\tau)|^2-|\delta\textbf{b}_{\perp}(t,\tau)|^2}{|\delta\textbf{v}_{\perp}(t,\tau)|^2+|\delta\textbf{b}_{\perp}(t,\tau)|^2}\right>,\label{eq:SRpAv}
\end{alignat}
The lowest range of frequencies measured here, $1/T_S < f < 1/T_A$, contains little variation in either $\left<\sigma_c\right>$ or $\left<\sigma_r\right>$. At frequencies $1/T_A < f < 1/T_O$, $\left<\sigma_c\right>$ increases and $\left<\sigma_r\right>$ decreases with increasing frequency. These trends stop at the outer scale $1/T_O$ and reverse at higher frequencies. For $\left<\sigma_r\right>$ this happens at $f \sim 1/T_O$ while for $\left<\sigma_c\right>$ there is an initial plateau before it starts decreasing at the highest frequencies. This high-frequency decrease in $\left<\sigma_c\right>$ is likely due to noise in the 3DP velocity observations \citep{Gogoberidze12, Podesta09, Chen13, Wicks13}.

\section{Analysis}

\begin{figure}[t]
\centering
\includegraphics[width=0.75\textwidth]{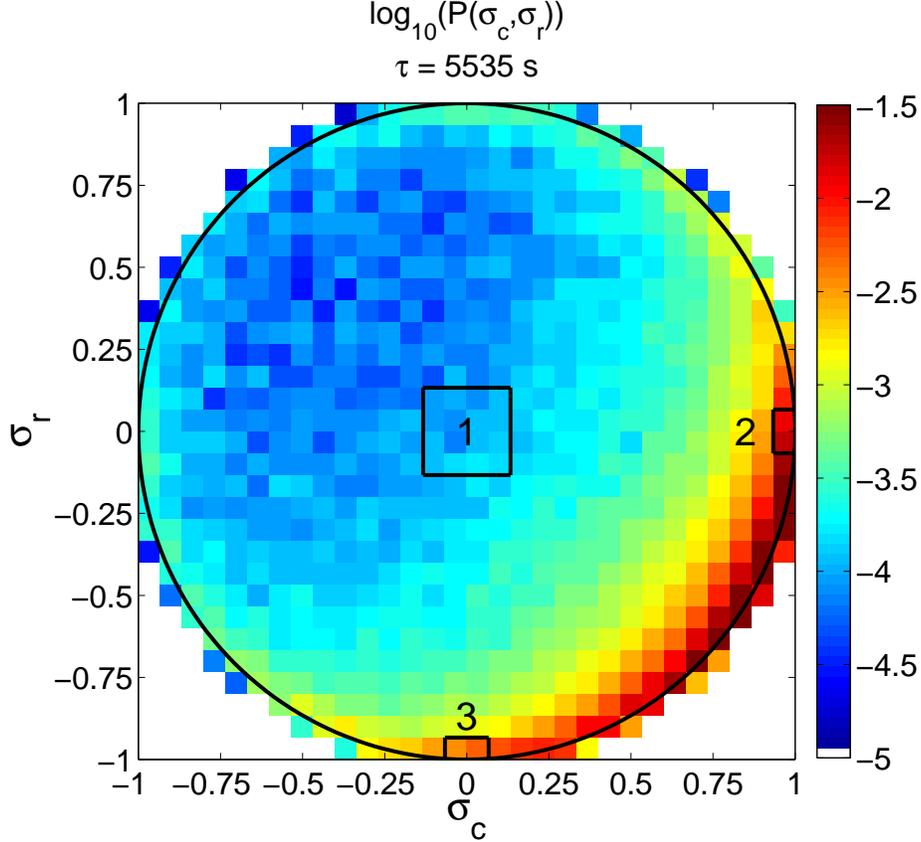}%
\caption{The joint probability distribution of $\sigma_c$ and $\sigma_r$ at time lag $\tau = 5535$~s for the entire 7-day analysis interval.} 
\label{fig:1}
\end{figure}

Having calculated the scale- and time-dependent $\sigma_c$ and $\sigma_r$, we can construct a scale-dependent joint probability distribution. The joint distribution at one of the scales in the $1/f$ range is shown in Figure \ref{fig:1}. Values of $\sigma_c$ and $\sigma_r$ must lie within a circle of radius $1$, as follows from their definitions (Equations \ref{eq:SCp}-\ref{eq:SRp}). The difference between fluctuations at the edge of this circle and in the middle is the geometry of the vectors relative to one another. Fluctuations at the edge of the parameter space are the most correlated. Indeed, at the edge of the circle, $\sigma_c^2 + \sigma_r^2 =1$, which implies
\begin{align}
|\delta\textbf{v}_{\perp}||\delta\textbf{b}_{\perp}| = & |\delta\textbf{v}_{\perp} \cdot \delta\textbf{b}_{\perp}|,\\
|\delta\textbf{z}^+_{\perp}||\delta\textbf{z}^-_{\perp}| = & |\delta\textbf{z}^+_{\perp} \cdot \delta\textbf{z}^-_{\perp}|,
\end{align}
so the Elsasser, velocity and magnetic-field fluctuations must be perfectly aligned (co-linear) at the edge.
\par
Close to the center of the circle, however $\sigma_c^2 + \sigma_r^2 \ll 1$, whence
\begin{equation}
|\delta\textbf{v}^2_{\perp} - \delta\textbf{b}^2_{\perp}| - 2|\delta\textbf{v}_{\perp} \cdot \delta\textbf{b}_{\perp}| \ll |\delta\textbf{v}^2_{\perp} + \delta\textbf{b}^2_{\perp}|, \label{eg:centerBV}
\end{equation}
\begin{equation}
|\delta\textbf{z}^{+2}_{\perp} - \delta\textbf{z}^{-2}_{\perp}| - 2|\delta\textbf{z}^+_{\perp} \cdot \delta\textbf{z}^-_{\perp}| \ll |\delta\textbf{z}^{+2}_{\perp} + \delta\textbf{z}^{-2}_{\perp}| \label{eg:centerPM}
\end{equation}
which can only be achieved when there are angles close to $90^{\circ}$ between the vectors in each equation (\ref{eg:centerBV} and \ref{eg:centerPM}) and the amplitudes of these vectors are approximately equal.
\par
Thus, by examining different regions of the $(\sigma_c$, $\sigma_r)$ space, we separates the two different types of correlations: in magnitude (equipartition) and direction (alignment).
\par
The probability distribution is strongly peaked along the edge of this parameter space where $\sigma_c > 0$ and $\sigma_r < 0$. This agrees well with the distributions found in other fast-wind intervals with different spacecraft by previous studies \citep{Bavassano98, Bavassano06, DAmicis10}. Here we extend the analysis to include a broader range of scales and to study the structure functions in different regions of this parameter space. Initially we concentrate on three regions of the $(\sigma_c$, $\sigma_r)$ space that have qualitatively distinct physical properties, these are shown as Regions 1, 2, and 3 in Figure \ref{fig:1} and their properties are summarized in Table \ref{table:1}. 
\par
\begin{table*}
\begin{tabularx}{\textwidth}{cccX}
Region & $\sigma_c$ & $\sigma_r$ & Description\\
 & & & \\
1 & $|\sigma_c| < 2/15$ & $|\sigma_r| < 2/15$ & Balanced ($\delta\textbf{z}^+_{\perp} \sim \delta\textbf{z}^-_{\perp}$) and equipartitioned ($\delta\textbf{v}_{\perp} \sim \delta\textbf{b}_{\perp}$) with unaligned vectors ($\cos(\phi) - \cos(\theta) \ll 1$). \\
2 & $\sigma_c > 14/15$  & $|\sigma_r| < 1/15$ & Strongly $\delta\textbf{z}^+_{\perp}$ dominated, equipartitioned and highly aligned ($\cos(\theta) \sim 1$).\\
3 &  $|\sigma_c| < 1/15$ & $\sigma_r < -14/15$ & Strongly $\delta\textbf{b}_{\perp}$ dominated, balanced and highly anti-aligned ($\cos(\phi) \sim -1$).
\end{tabularx}
\caption{Properties of the three regions used in Figures \ref{fig:1} and \ref{fig:2}. The angles between vectors are defined as $\cos(\phi) = \delta\textbf{z}^+_{\perp}\cdot\delta\textbf{z}^-_{\perp}/|\delta\textbf{z}^+_{\perp}||\delta\textbf{z}^-_{\perp}|$ and $\cos(\theta) = \delta\textbf{v}_{\perp}\cdot\delta\textbf{b}_{\perp}/|\delta\textbf{v}_{\perp}||\delta\textbf{b}_{\perp}|$.}
\label{table:1}
\end{table*}
Region 1 is in the center of the parameter space where $|\sigma_c| < 2/15$ and  $|\sigma_r| < 2/15$. When $\sigma_c \ll 1$ and $\sigma_r \ll 1$ fluctuations can be described as balanced ($\delta\textbf{z}_{\perp}^+ \sim  \delta\textbf{z}_{\perp}^-$) and Alfv\'{e}nically equipartitioned ($\delta\textbf{b}_{\perp} \sim \delta\textbf{v}_{\perp}$) and as a result are unaligned, with the cosine of the angle between $\delta\textbf{v}_{\perp}$ and $\delta\textbf{b}_{\perp}$ $|\cos(\theta)| < 2/15$ and the cosine of the angle between $\delta\textbf{z}^+_{\perp}$ and $\delta\textbf{z}^-_{\perp}$ $|\cos(\phi)| < 2/15$.
\par
Region 2 contains fluctuations with $\sigma_c > 14/15$ and $|\sigma_r| < 1/15$, consistent with very pure outward propagating Elsasser fluctuations. These are, therefore, imbalanced ($\delta\textbf{z}_{\perp}^+ \gg  \delta\textbf{z}_{\perp}^-$), but equipartitioned and aligned. 
\par
Region 3 has $|\sigma_c| < 1/15$ and $\sigma_r < -14/15$, meaning that the fluctuations in it are magnetically dominated and, therefore, balanced and have antialigned Elsasser fields. 
\par
These values are chosen so that the probability of fluctuations being observed does not change systematically across each box but that there are enough ($> 30$) observations in the box at each scale $\tau$ to calculate accurate structure-function averages. The regions must also be symmetrical about whichever variable is close to 0 in order to make fluctuations balanced (Region 3), equipartitioned (Region 2), or balanced and equipartitioned (Region 1) on average.
\par
We now calculate the structure functions only using those fluctuations that have $\sigma_c$ and $\sigma_r$ corresponding to one of these three regions. These are scale-dependent structure functions conditioned on the local correlation properties of the fluctuations. Thus these conditioned structure functions do not necessarily come from continuous sections of the time series, but are aggregated from separate times across the whole time series. To do this we make the assumption that the time series of fluctuations are stationary so that fluctuations that are not locally neighboring may still be statistically comparable. This is a reasonable assumption to make for this particular 7-day interval because the data show little systematic variation in magnetic-field strength or proton density. There is a trend of decreasing solar wind speed with time during the interval, however the instantaneous speed $|V|$ remains within one Alfv\'{e}n speed of the average solar wind speed $\left<|V|\right>$ (i.e. $|V|$ is always found within the range $\left<|V|\right> \pm V_A$) over the entire interval. Thus the Alfv\'{e}n Mach number, Reynolds number and other related quantities do not change significantly over the interval. 
\par
\begin{figure}[t]
\centering
\includegraphics[width=0.75\textwidth]{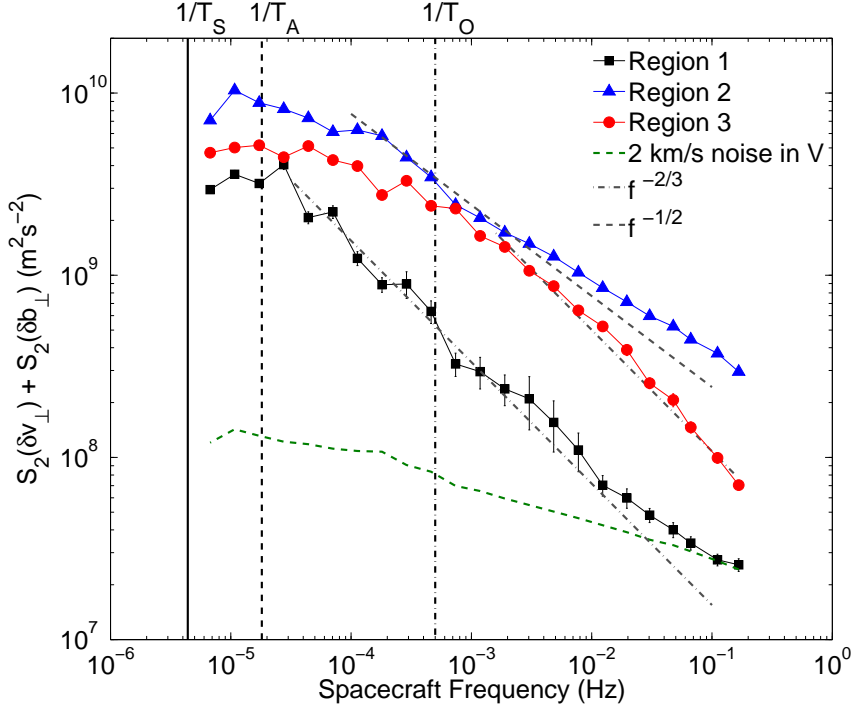}%
\caption{The sum of the velocity and magnetic field structure functions in three regions of the $\sigma_c,\sigma_r$ plane: Region 1, which contains balanced, equipartitioned fluctuations, Region 2, which contains very pure anti-sunward Alfv\'{e}nic fluctuations, and Region 3, which contains very pure magnetic fluctuations.}
\label{fig:2}
\end{figure}

We compare the structure functions from the three separate regions in Figure \ref{fig:1} by plotting the sum of $S_2(\delta\textbf{v}_{\perp})$ and $S_2(\delta\textbf{b}_{\perp})$ against frequency in Figure \ref{fig:2} . We can see that the balanced, equipartitioned and unaligned fluctuations taken from Region 1 scale steeply from close to the large-scale limit of the turbulence at $f \sim 1/T_A$ down to the instrument noise floor at small scales (shown by the dashed green line; see \cite{Podesta09, Gogoberidze12, Wicks13}). The scaling of these fluctuations is close to $f^{-2/3}$ and therefore suggests active nonlinear interaction. There is no discernible ``spectral break" at the ``outer scale" $1/T_O$. The structure functions of the fluctuations from regions 2 and 3 have flatter scaling in the low-frequency regime and then steepen at higher frequencies. Region 3 (magnetically dominated fluctuations) shows the same steep scaling as region 1, albeit starting at a much higher frequency comparable to the outer scale $1/T_O$ originally estimated from the full data set. Region 2 (imbalanced Elsasser fluctuations) never reaches a scaling as steep as $f^{-2/3}$ (a $-5/3$ spectrum) and appears, in the inertial range, to have scaling that is somewhat shallower even than $f^{-1/2}$ (a $-3/2$ spectrum), although this may be affected by instrument noise \citep{Gogoberidze12, Podesta09, Wicks13}. 
\par
Having ascertained that the scaling of the conditional structure functions is distinct in three distinct regions of the ($\sigma_c$, $\sigma_r$) space, even within the single fast stream that these data originated from, we can go further with our analysis. We now construct structure functions conditioned on each individual pair of values $\sigma_c$ and $\sigma_r$ represented by the pixels in Figure \ref{fig:1}. The pixels are arranged in a square $30\times30$ grid across the space $|\sigma_c| < 1$, $|\sigma_r| < 1$. We remove the pixels that have fewer than 30 observations at each scale to filter out the least well determined structure functions. The average total fluctuation amplitude $S_2(\delta\textbf{v}_{\perp}, \tau) + S_2(\delta\textbf{b}_{\perp}, \tau)$ in each pixel is plotted and the exponent of the summed structure functions corresponding to each pixel is measured by fitting a straight line to the amplitudes of five neighboring scales on a log-log scale.
\par
\begin{figure*}
\centering
\includegraphics[trim = 10mm 0mm 40mm 0mm, clip, width=\textwidth]{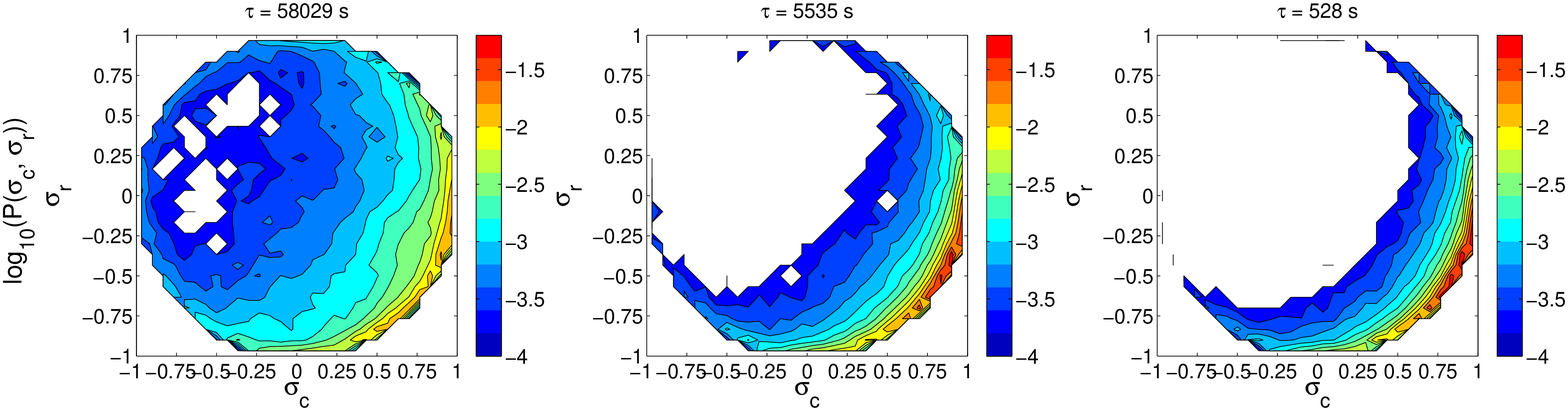} \\%
\includegraphics[trim = 10mm 0mm 40mm 0mm, clip, width=\textwidth]{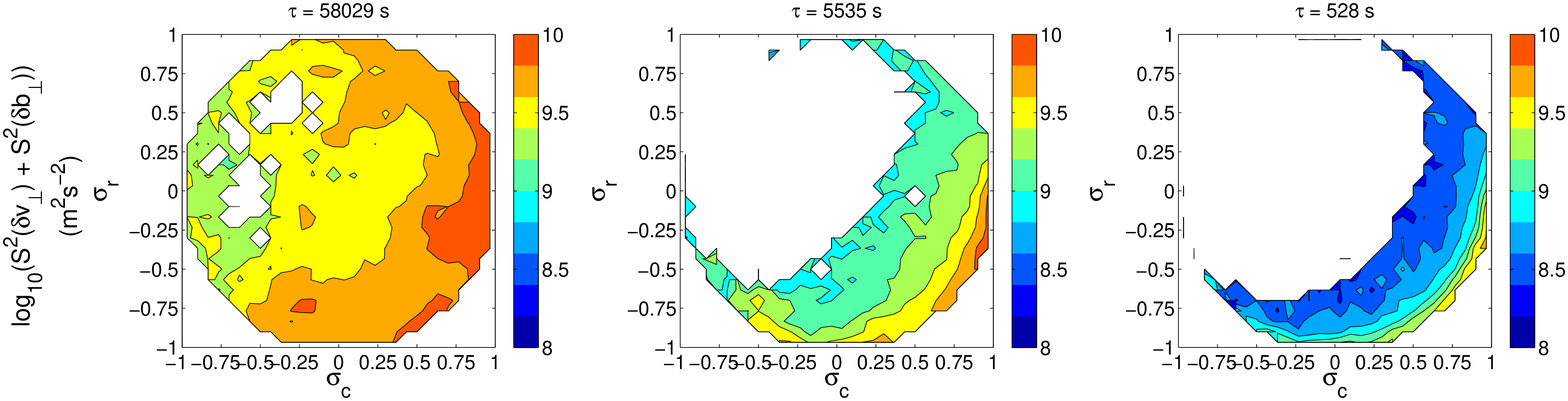} \\%
\includegraphics[trim = 10mm 0mm 40mm 0mm, clip, width=\textwidth]{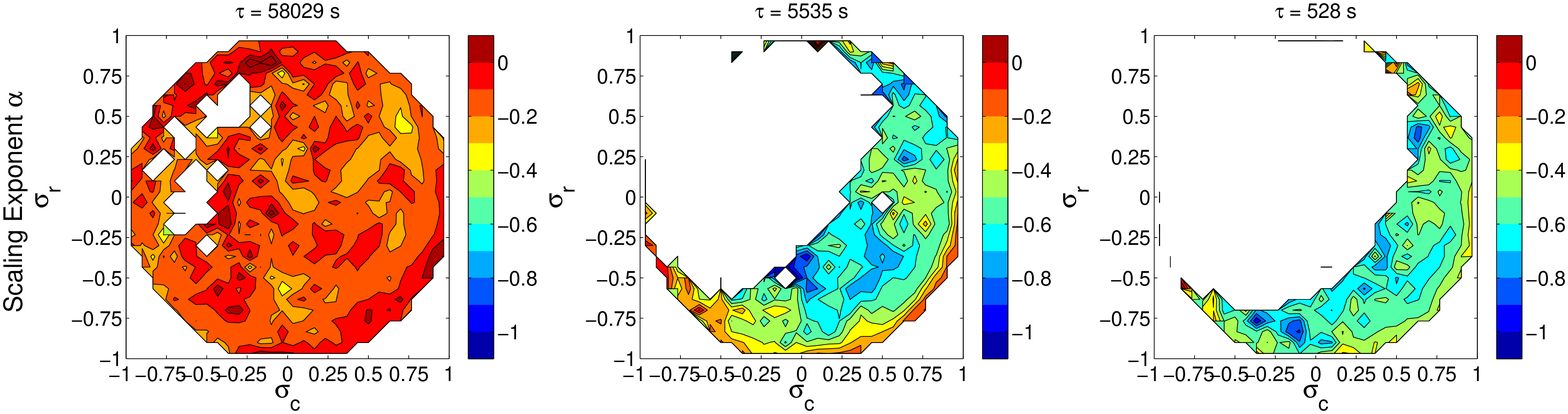}\\%
\caption{The evolution of different properties of structure functions from the Alfv\'{e}n interaction time $T_A$, through the $1/f$ range and into the inertial range plotted in the ($\sigma_c, \sigma_r$) plane. The top row shows the joint probability distribution of $\sigma_c$ and $\sigma_r$. The second row shows the total amplitude in fluctuations, $S_2(\delta\textbf{v}_{\perp}) + S_2(\delta\textbf{b}_{\perp})$. The third row shows the scaling exponent of the structure functions, $S_2 \propto f^{\alpha}$, measured over five consecutive points centered on the scale $\tau$ of each panel. The three columns correspond to the three time scales indicated by arrows in Figures \ref{fig:0.1} and \ref{fig:2}. The middle row is for the same $\tau$ as Figure \ref{fig:1}.}
\label{fig:3}
\end{figure*}

The top row of panels in Figure \ref{fig:3} shows the joint probability distribution of $\sigma_c$ and $\sigma_r$, the middle row shows the sum of velocity and magnetic field structure functions at each scale, giving an estimate of the total energy in the fluid motion, and the bottom row shows the structure function exponent. These are each measured at three different scales covering three decades in frequency. These scales are $\tau = 58029$~s $\sim T_A$, with the exponent measured in the range $5\times10^{-6} < f < 5\times10^{-5}$~Hz (left column), the middle of the $1/f$ range at $\tau = 5535$~s, with the exponent measured in the range $5\times10^{-5} < f < 5\times10^{-4}$~Hz (middle column), and the top of the inertial range at $\tau = 528$~s, with the exponent measured in the range $5\times10^{-4} < f < 5\times10^{-3}$~Hz (right column). These scales are indicated by arrows in Figures \ref{fig:0.1} and \ref{fig:2}.
\par
Moving from left to right along the top row of Figure \ref{fig:3} we see how the joint probability distribution of $\sigma_c$ and $\sigma_r$ changes through the $1/f$ range. The probability is always higher in the bottom right corner (imbalanced fluctuations with some excess of magnetic energy), as was seen in Figure \ref{fig:1}, but the peak becomes more pronounced as the scale $\tau$ becomes smaller. The middle row of panels shows the corresponding fluctuation amplitudes. At the largest scale (left panel), the energy is predominantly in $\sigma_c > 0$ outward propagating Elsasser fluctuations, but this maximum is larger than the minimum by less than an order of magnitude. In the inertial range (right panel) the energy has been concentrated into a narrow band along the bottom right edge of the parameter space ($\sigma_c > 0, \sigma_r < 0$), which now has approximately two orders of magnitude more power, on average, than fluctuations found closer to the center of the circle.
\par
The third row of panels show the structure-function scaling exponent. At the largest scales (left panel) the exponent is close to $0$ everywhere. In the middle panel, which represents most of the $1/f$ range, the bottom edge of the ($\sigma_c, \sigma_r$) space has a flat exponent but large areas of the space closer to the middle have instances with steeper scalings in the range $-0.4 < \alpha < -0.8$, characteristic of active nonlinear interactions. The flattest values occur in the range $(-1 < \sigma_c < -0.5, -0.75 < \sigma_r < -0.25)$, corresponding to anti-aligned $\delta\textbf{v}_{\perp}$ and $\delta\textbf{b}_{\perp}$ as well as anti-aligned Elsasser fluctuations with a dominant $\delta\textbf{b}_{\perp}$ component. At the lowest inertial range frequencies (right panel) the exponent has steepened to $\alpha < -0.4$ almost everywhere, although the gradients at the edge of the ($\sigma_c, \sigma_r$) space typically remain flatter than those closer to the center.
\par
Figures \ref{fig:2} and \ref{fig:3} show that the local correlation properties of velocity and magnetic field fluctuations in the solar wind have a strong effect on the scale at which the onset of turbulence occurs. Those fluctuations with the widest $1/f$ range are dominated by magnetic fluctuations ($\sigma_r \approx -1$) or outward Elsasser fluctuations ($\sigma_c \approx 1$) or a mixture of both. Magnetic field fluctuations without an associated velocity fluctuation are likely to be force-free structures (where $\textbf{j}\times\textbf{B} = 0$), and so a stable solution to the MHD equations. Unidirectional Alfv\'{e}n-wave packets are also a stable solution to the MHD equations (Elsasser solutions). Thus these two regions of the joint probability distribution of $\sigma_c$ and $\sigma_r$, or regions dominated by a mixture of the two, most likely to be found at the bottom-right edge of the distribution, may be more stable to nonlinear interaction than other regions of the distribution.

\section{Conclusions}

This analysis has shown that low-frequency fluctuations in the solar wind can be meaningfully organized according to the values of their normalized cross helicity $\sigma_c$ and residual energy $\sigma_r$ \citep{Bavassano98, Bavassano06, DAmicis10}. Most fluctuations cluster near the edge of the circle $\sigma_c^2 + \sigma_r^2 = 1$, with $\sigma_c > 0$ and $\sigma_r < 0$. This means that they are a mixture of imbalanced fluctuations dominated by the outward propagating Elsasser field and magnetically dominated structures. These two types of fluctuations, in their purest form, are also exact nonlinear solutions of the MHD equations (Elsasser and force-free solutions, respectively) and so it stands to reason that they would be the most resilient ones to survive nonlinear interactions leading to a turbulent cascade. Indeed, we find that the width of the ``non-turbulent" $1/f$ range is largest for these fluctuations. In contrast, the subdominant fluctuations with small $\sigma_c$ and $\sigma_r$ (Elsasser- and Alfv\'{e}nically balanced ones) exhibit a robust Kolmogorov-like scaling starting deep inside what is normally viewed as the $1/f$ range and showing no spectral break at the usual value of the ``outer scale" (set by the imbalanced and magnetically dominated fluctuations).
\par
Observations showing scale-dependent alignment of velocity and magnetic field fluctuations at low frequencies \citep{Podesta09, Hnat11} and Elsasser fluctuations \citep{Wicks13} can be understood in the context of these results. Unaligned fluctuations of both types (found away from the edge of the circular $\sigma_c$ and $\sigma_r$ space) are removed by selective nonlinear interaction, which preserves the more aligned ones (closer to the edge). Note, however, that this behavior is not an automatic consequence of the fact that nonlinear interactions are stronger for the balanced, unaligned, Alfv\'{e}nic fluctuations: there is no separate conservation law for these, so they are not obliged to cascade into similarly balanced, unaligned, Alfv\'{e}nic fluctuations at smaller scales. It is thus quite interesting that when selected by conditioning on the values of $\sigma_c$ and $\sigma_r$, they give rise to what appears to be quite a robust Kolmogorov scaling. A complete theory of MHD turbulence should strive to explain this behavior (which might perhaps provide a valuable hint).
\par
Since cross helicity is typically correlated with solar wind speed, source region and distance from the Sun, our results are consistent with previous observations showing that slower, less correlated streams have smaller $1/f$ ranges and faster, more correlated streams have larger $1/f$ ranges \citep{Tu89, Tu90}. As the solar wind travels further out into the heliosphere, the cross helicity decreases and so does the $1/f$ range \citep{Bavassano82, Roberts10, Tu90}, again agreeing qualitatively with the conclusions presented here that less correlated fluctuations have narrower $1/f$ ranges. Further work is required to prove the universality of our results and to investigate the effect of source region and radial evolution on the joint probability distributions of $\sigma_c$ and $\sigma_r$. Numerical simulations of MHD turbulence could also be used to investigate the link between the geometry of the fluctuations as defined by $\sigma_c$ and $\sigma_r$ and the strength of the turbulent nonlinear interaction.
\par
An improvement to the method used here would be to include non-Gaussian particle distribution effects, such as the pressure anisotropy and beam drift speed, in the Alfv\'{e}n normalization for the magnetic field \citep{Chen13}. Using this improved normalization for the analysis presented here decreases the average $\sigma_r$, but the quantitative and qualitative results showing the existence of wide $1/f$ ranges at the edge of the joint probability distribution of $\sigma_c$ and $\sigma_r$, but not near the center, remain true. 

\section{Acknowledgments.}

This research was supported by the NASA Postdoctoral Program at the Goddard Space Flight Center (RTW), STFC (AM, TSH), NASA contracts NNN06AA01C and NAS5-02099 (CHKC) and the Leverhulme Trust Network for Magnetized Plasma Turbulence. Wind data were obtained from the SPDF website http://spdf.gsfc.nasa.gov.

\end{document}